\newcommand{\beq}{\begin{equation}}
\newcommand{\eeq}{\end{equation}}
\newcommand{\bea}{\begin{eqnarray}}
\newcommand{\eea}{\end{eqnarray}}
\newcommand{\bear}{\begin{eqnarray*}}
\newcommand{\eear}{\end{eqnarray*}}

\newcommand{\rf}[1]{(\ref{#1})}
\documentstyle[aps,preprint]{revtex}
\begin{document}

\draft

\title
{EXACT SOLUTION OF THE ASYMMETRIC EXCLUSION MODEL WITH PARTICLES
 OF ARBITRARY SIZE}

\author{
F. C. Alcaraz}  
\address{Departamento de F\'{\i}sica, 
Universidade Federal de S\~ao Carlos, 13565-905, S\~ao Carlos, SP
Brazil}
\author{ R. Z. Bariev}

\address{Departamento de F\'{\i}sica, 
Universidade Federal de S\~ao Carlos, 13565-905, S\~ao Carlos, SP
Brazil}

\address{The Kazan Physico-Technical Institute of the Russian 
Academy of Sciences, 
Kazan 420029, Russia}

\maketitle

\begin{abstract}

	A generalization of the simple exclusion asymmetric model is 
introduced. In this model an arbitrary mixture of molecules with 
distinct sizes $s = 0,1,2,\ldots$, in units of lattice space, diffuses 
asymmetrically 
on the lattice. A related surface growth model is also presented. 
Variations of the distribution of molecules's sizes may change the 
excluded volume almost continuously. We solve the model exactly 
through the Bethe ansatz and the dynamical critical exponent $z$ is 
calculated from the finite-size corrections of the mass gap of the 
related quantum chain. Our results show that for an arbitrary distribution 
of molecules the dynamical critical behavior is on the Kardar-Parizi-Zhang 
(KPZ) universality. \\
	{ \bf To appear in Phys. Rev. E (1999).}

\end{abstract}


\narrowtext    
\section{Introduction}

The asymmetric simple exclusion model \cite{ligget} 
 is a one-dimensional stochastic model
that describes the time fluctuations of particles diffusing asymmetrically 
on the lattice. If we interprete an occupied site as $\sigma_i^z = +1$ and a
vacant site as $\sigma_i^z = -1$ the time evolution of the probability 
distribution of particles is given by the following asymmetric XXZ Hamiltonian 
\begin{equation}
\label{1}
H = -\sum_{i=1}^N\lbrack\epsilon_+\sigma_j^-\sigma_{j+1}^+ +\epsilon_-
\sigma_j^+\sigma_{j+1}^- + \frac{1}{4}(1-\sigma_j^z\sigma_{j+1}^z)\rbrack, 
\end{equation}
where $N$ is the number of lattice sites and $\sigma_i^{\pm} =  
(\sigma^x \pm i\sigma^y)/2$ are the raising and lowering Pauli 
operators. Periodic boundary conditions  are imposed and $\epsilon_+$
and $\epsilon_-$  ($\epsilon_+ + \epsilon_- = 1$) are the transition 
probabilities for having a motion to the right and to the left,  respectively.
 The physical properties of this non-hermitian quantum chain  
as well as the related asymmetric six-vertex model are still under 
extensive investigations \cite{kim2,albertini1,albertini2}.
This model also describes the surface fluctuations in a growth model
known as single-step model \cite{meakin,liu} where  
$(\sigma_i^z + \sigma_{i+1}^z)/2 = -1, 0, 1 $
is the height difference between nearest-neighbor steps located at 
the odd-half integers sites $(i-1/2)$ ($i=1,2,\ldots$).
The master equation defining the Hamiltonian \rf{1} can also be interpreted 
\cite{spohn}
as the discretized version of the noisy Burges equation or the 
Kardar-Parisi-Zhang (KPZ) equation \cite{KPZ} 
governing the motion of the height 
of growing 
surfaces whose local growth velocity depends nonlinearly on the local shape.
The conection between the scaling behavior of the structure function 
\cite{liu,krug} of the 
stochastic model and the finite-size dependence  of the real part of the mass 
gap $G_N$   gives us the dynamical critical exponent $z$,
\bea
\mbox{Re}(G_N) \sim N^{-z}.\nonumber
\eea
This connection enabled Gwa and Spohn \cite{spohn} to explore the Bethe ansatz 
solution of 
\rf{1} and calculate exactly the exponent $z=3/2$ in the highly anisotropic 
limit $\epsilon_- = 0$. Subsequently Kim \cite{kim} extended  this result for
$\epsilon_- > 0 $.

In a previous paper \cite{alcabar1} we observed, in connection with a model 
for strongly correlated system, 
that it is possible to keep the exact integrability of the XXZ chain by
enlarging the excluded volume to the spins. Motivated by these results we 
introduce in this paper a generalization of the simple exclusion model 
where each particle individually instead of having size 1, in units of 
lattice spacing, may have an arbitrary and distinct integer size 
$(s_i = 0,1,2,\ldots)$. 
Particles with size $s_i > 1 $ will produce stronger excluded volume 
 than those in the simple exclusion 
model where all the particles have unity size $(s_1 = s_2 =\cdots=s_n = 1)$. 
Particles with size zero $(s = 0)$
produce  no excluded  volume  since we may put an arbitrary number 
of them  at the same site. By considering 
arbitrary mixtures of molecules with appropriate sizes we may 
change continuously the excluded volume  in the bulk limit.
The time-evolution operator for these models 
are generalizations of ferromagnetic XXZ chain where 
restrictions on spin configurations, which depend on the particular
sizes of the molucules $s_i$, are added. We show that for arbitrary
distribution  of molecule's sizes the eigenspectrum of the 
related Hamiltonian can be calculated exactly through the Bethe 
ansatz. The exact integrability for the particular case where all 
the molecules have the same size and the diffusion is fully asymmetric 
was also verified recently by 
Sasamoto and Wadati \cite{sasawada}.

Following Gwa and Spohn \cite{spohn} we show, in the anisotropic limit 
$\epsilon_- = 0$, that for arbitrary distribution of molecules the 
real part of the gap behaves as $\mbox{Re}(G) \sim N^{-3/2}$, giving a universal
KPZ behaviour with dynamical critical exponent $z=3/2$. Since in our model
the excluded volume  can be controlled continuously by 
changing the distribution of molecules, the above exact results imply
that the exclusion volume effect is irrelevant to the KPZ dynamics.

The paper is organized as follows. In the next section we introduce the 
generalized asymmetric model and the related generalized surface growth 
model.  The Bethe-ansatz solution  of our model is presented in section 3 and 
in section 4 a   
 numerical and analitical calculation of the dynamical critical 
exponent $z$ is presented.  Finally in section 5 we give our 
conclusions and in an appendix we relate exactly 
the general asymmetric exclusion model with several boundary conditions with 
the simple exclusion model, in different lattice sizes.

\section{The generalized asymmetric exclusion model}

The simplest realization of the model we consider in this paper is 
the asymmetric diffusion of molecules (or particles)  on a lattice of size $N$,
where each  molecules $i$ ($i=1,2,\ldots,n$)  may have 
a distinct size $s_i = 0,1,2,\ldots$, in units
of lattice spacing. 
We represent the molecules on the lattice by placing their center of mass 
at the lattice sites. In Fig. 1 we show some examples of configurations 
of $n=5$ molecules with size's distribution $\{s\}$ in a lattice with $N=6$ 
sites.  
Molecules  of size $s = 0$ are special since 
 we can put in a given lattice point an 
arbitrary number of them.  As we can see from Fig. 1,  the minimum distance 
between two particles with size $s,s'$ on the lattice is given by 
\begin{equation}
\label{1-1}
d_{s,s'} = \mbox{Int}\lbrack\frac{s + s' + 1}{2}\rbrack, \;\;\;\; 
s,s' = 0,1,2,..., 
\end{equation}
where $\mbox{Int}(x)$  is the integer part of $x$.
 In order to describe the occupancy of a given 
site $i$ ($i=1,2,\ldots,N$) 
we attach on it a site variable $\beta_i$ taking integer values 
$(\beta_i \in {\rm Z})$. If $\beta_i = 0$ the site is empty, on the other hand 
if $\beta_i > 0$ we have a molecule of size $s=\beta_i$ and if 
$\beta_i = -n < 0$ we have $n$ molecules of size 0. The allowed 
configurations $\lbrace\beta_i\rbrace = \lbrace\beta_1,\beta_2,...,
\beta_N\rbrace$ are those satisfying the constraints imposed by the 
sizes of molecules in a periodic lattice, i.e., if $\beta_i \ne 0$ and
$\beta_j \ne 0$ we must have  $|i-j|\ge d_{s(\beta_i),s(\beta_j)}$, where
$s(\beta) = 0$ if $\beta 
\leq 0$, and $s(\beta) = \beta$ if $\beta > 0$, is the excluded volume 
 (or size) associated to $\beta$.

The master equation for the probability distribution 
$P(\lbrace\beta\rbrace,t)$ can be written in general as
\beq
\label{2} 
\frac{\partial P(\lbrace\beta\rbrace,t)}{\partial t} = - \Gamma(\lbrace\beta\rbrace 
\rightarrow \lbrace\beta'\rbrace) P(\lbrace\beta\rbrace,t) +
\Gamma(\lbrace\beta'\rbrace 
\rightarrow \lbrace\beta\rbrace) P(\lbrace\beta'\rbrace,t),
\eeq
where $\Gamma(\lbrace\beta\rbrace \rightarrow \lbrace\beta'\rbrace)$ is 
the transition rate, where a configuration $\lbrace\beta\rbrace$ changes to
$\lbrace\beta'\rbrace$.  In the model under  consideration we only allow,
whenever it is possible, a single particle to diffuse into its nearest 
neighbor sites. The possible motions are diffusion to the right
\bea
\label{3}
\beta_i\;\;\emptyset_{i+1} &\rightarrow& \emptyset_i\;\;\beta_{i+1},\;\;\;\;\;\;\;\;
\;\;\;\;\;\;\;\beta > 0\nonumber\\
\beta_i\;\;\gamma_{i+1} &\rightarrow& (\beta+1)_i \;\;(\gamma - 1)_{i+1}
,\;\;\;\;\beta < 0,
\gamma \leq 0,
\eea
with transition rate $\epsilon_R$ and diffusion to the left
\bea
\label{4}
\emptyset_{i}\;\;\beta_{i+1} &\rightarrow&\beta_i\;\;\emptyset_{i+1},\;\;\;
\beta > 0\nonumber\\
\gamma_i\;\;\beta_{i+1} &\rightarrow& (\gamma -1)_i\;\;(\beta+1)_{i+1}
 \;\;\;\beta < 0, \;\;\;\; \gamma \leq 0, 
\eea
with transition rate $\epsilon_L$. The master equation (2) can be written as 
a Schr\"odinger equation in Euclidean time (see ref.\cite{alcrit1} for general
application for two-body processes)
\beq
\label{5}
\frac{\partial|P>}{\partial t} = -H |P>,
\eeq
if we interpret $|P> \equiv P(\lbrace\beta\rbrace, t)$ as the associated wave 
function. If we represent $\beta_i$ as $|\beta>_i$ the vector 
$|\beta>_1\otimes |\beta>_2 \otimes \cdots \otimes |\beta>_N$ will give us 
the associated 
Hilbert space. The process \rf{3} and \rf{4} gives us the Hamiltonian
\bea
\label{6}
H &=& -D \; P\sum_{i=1}^N (H_{i,i+1}^{>} + H_{i,i+1}^{<})P, \nonumber \\
H_{i,j}^{>}& =& \sum_{\beta =1}^{\infty}\lbrack \epsilon_{+}
(E_i^{0,\beta}E_{j}^{\beta,0}- E_{i}^{\beta,\beta}E_{j}^{0,0}) +
\epsilon_{-}(E_i^{\beta,0}E_{j}^{0,\beta }- E_{i}^{0,0} E_{j}^{\beta,\beta})
\rbrack , \\
H_{i,j}^{<}& =& \sum_{\beta =-\infty}^{-1}\sum_{\gamma =-\infty}^{0}\lbrack 
\epsilon_{+}
(E_i^{\beta +1,\beta}E_{j}^{\gamma -1,\gamma}- E_{i}^{\beta,\beta}E_{j}^
{\gamma,\gamma}) +
\epsilon_{-}(E_i^{\gamma-1,\gamma}E_{j}^{\beta +1,\beta }- 
E_{i}^{\gamma,\gamma} E_{j}^{\beta,\beta})
\rbrack \nonumber ,
\eea
with 
\beq
\label{7}
D = \epsilon_R + \epsilon_L,\;\;\; \epsilon_+ =\frac{\epsilon_R}{\epsilon_R + 
\epsilon_L},\;\;\; \epsilon_- = \frac{\epsilon_L}{\epsilon_R + \epsilon_L},
\eeq
and periodic boundary conditions. The matrices $E^{\alpha,\beta}$ are 
infinite-dimensional with a single non-zero
element $(E^{\alpha,\beta})_{i,j} = \delta_{\alpha,i}\delta_{\beta,j}
(\alpha,\beta,i,j \in {\rm Z})$. The projector P keeps on the Hilbert 
space only the vectors $|\lbrace\beta\rbrace>$ 
satisfying the constraint \rf{1-1} which mathematically means that 
 for  the all  
$\beta_i,\beta_j \ne 0,\;\; |i - j| \ge  d_{s(\beta_i),s(\beta_j)}$.
The constant D in  \rf{6} fixes the time scale and for simplicity we
 chose $D = 1$. A particular simplification of the above Hamiltonian
occurs when all the molecules have the same fixed size $s > 0$. In this case
the Hamiltonian can be expressed in terms of spin-$\frac{1}{2}$ Pauli
matrices
\beq
\label{8}
H_{\lbrace s_1=\cdots=s_n = s\rbrace} = - \; 
P_s\{\sum_{i=1}^N\lbrack \epsilon_+
\sigma_i^-\sigma_{i+1}^+ +\epsilon_-\sigma_i^+\sigma_{i+1}^- +
\frac{1}{2}(\epsilon_+ + \epsilon_-)(\sigma_i^z\sigma_{i+1}^z - 1)\rbrack\} P_s,
\eeq
where now $P_s$  projects out configurations where two up spins, in 
$ \sigma^z$ basis, are at distance smaller than the size $s > 0$. In the
case where $s=1$ the projector $P_s = 1$ and we have the standard asymmetric 
exclusion Hamiltonian \rf{1}. In terms of Pauli matrices this operator 
has the general form
\beq
\label{9}
P_s = \prod_{i}\Bigl[\frac{1}{2}(1 - \sigma_i^z) + 
\frac{1}{2}(1 + \sigma_i^z)
\prod_{l=1}^{s-1}(\frac{1 - \sigma_{i+l}^z}{2})\Bigr].
\eeq
The Hamiltonian \rf{8} can be more easily compared with standard magnetic 
quantum chains by performing for $\epsilon_+,\epsilon_- \ne 0$ the following 
canonical change of variables
\beq
\label{10}
\sigma_i^{\pm} \rightarrow (\frac{\epsilon_-}{\epsilon_+})^{\pm i/2}
\sigma_i^{\pm},\;\;\; \sigma^z \rightarrow \sigma^z,\;\;\; i = 1,2,...,N,
\eeq
which gives 
\bea
\label{11}
H &=& -\frac{1}{2}\sqrt{\epsilon_+\epsilon_-}\sum_{i=1}^N P_s\Bigl[
\sigma_i^x\sigma_{i+1}^x
+ \sigma_i^y\sigma_{i+1}^y + \Delta (\sigma_i^z\sigma_{i+1}^z-1)\Bigr] 
P_s,\nonumber\\
\Delta &=&\frac{\epsilon_+ + \epsilon_-}{\sqrt{\epsilon_R \epsilon_L}}.
\eea
Apart from the projector $P_s$ this is the ferromagnetic XXZ chain or 
the anisotropic Heisenberg chain. However in distinction with (8) the boundary
conditions are now twisted
\beq
\label{12}
\sigma_{N+1}^{\pm} = (\frac{\epsilon_+}{\epsilon_-})^{\pm N/2}
\sigma_1^{\pm},\;\;\; \sigma_{N+1}^z =\sigma_1^z.
\eeq
We expect that ferromagnetic quantum chains like those in \rf{11} are gapped 
for $\Delta > 1$ However since 
$(\epsilon_+/\epsilon_-)^{N/2}\rightarrow \infty$, for 
$N\rightarrow \infty$ the boundary condition gives us interaction with the 
same degree of importance as the totality of the other interactions (see 
\cite{henkel} for a related problem). As we 
will see, from the exact solution of \rf{6} and \rf{8}, this surface term is 
strong enough  to produce a gapless eigenspectrum.

In surface growth physics the asymmetric 
simple exclusion model is related 
 to the single-step model. Similarly our generalized model is also 
related to a generalization of the single-step model. The surface 
configurations in this growth model are obtained by defining 
height variables $h_i$ $(i =1,2,\ldots )$ which are related to the 
spin variables in 
our generalized asymmetric diffusion model. For simplicity we are going to 
present only the   
  surface growth model related to the diffusion problem where  
all molecules have the 
same size $s$. Let us consider initially $s > 0$. For a given 
configuration $\lbrace\beta_1,...,\beta_N\rbrace$ of molecules of size $s$,
the height variables should obey
\beq
\label{f1}
h_{i+1} - h_i = f(\beta_{i -\frac{s}{2}}, \beta_{i -\frac{s}{2}+1},...,
\beta_{i +\frac{s}{2}}, \beta_{i +\frac{s}{2}+1}),
\eeq
where $f = 0$ for all allowed configurations except in the case 
\beq
\label{f2}
f(0,\ldots,0) = -1\;\;\;\; \mbox{and}\;\;\;\; f(1,0,\ldots,0,1) = 1.
\eeq
The variables $\lbrace\beta\rbrace$ of the related diffusion model are defined 
at the links or at the same 
positions of the height variables $\lbrace h \rbrace$
depending if the size of the molecules $s$ is odd or even, respectively.
The number of molecules $n$ $(0,1,\ldots)$, 
in the generalized asymmetric diffusion 
is 
conserved and for each value of $n$ we are going to have a growth model with 
different boundary 
conditions in the spacial direction. The dynamical
rules defining the growth are the following 

a) No steps on the surface are 
allowed to be higger than 1, in units of lattice spacing in the growth 
direction, i. e., 
\beq
\label{f3}
h_{i+1} -h_i = 1, 0,-1 \quad (i =1,\ldots,n-1).
\eeq

b) All the local valleys and hills should have a size, in units 
of the lattice spacing in the spatial direction, which is a multiple of
$b = s+1$.

c) The following boundary condition should be satisfied 
\beq
\label{f4}
h_{i+N} = h_i - \bar{h}, \;\;\;\;\bar{h} = b\lbrace[N/b]_I - n\rbrace
+ [N/b]_R  .
\eeq
where by $[N/b]_I$ and $[N/b]_R$ we mean the integer part and the rest 
of the division $N/b$.

d) The surface changes whenever obeying the previous requirements we 
can still adsorb $ (h_{i+l} \rightarrow h_{i+l}+1, l = 0,1,...,b-1)$
or desorb $(h_{i+l}\rightarrow h_{i+l}-1, l =0,1,...,b-1)$ at height 
$h_i$ $(i=1,\ldots,N)$ a retangular brick of size $b$ in the spatial 
direction and size 1 in the growth direction. 

We choose a height $h_i$ $ (i = 
1,...,n)$ at random. If it is possible to adsorb or desorb a brick, 
with probability $\epsilon_+/2$ $ (\epsilon_-/2)$ we desorb (adsorb) 
a brick, and do nothing with probability $\frac{1}{2}$. If it is possible
, at $h_i$, only to desorb (adsorb) a brick, with probability 
$\epsilon_+$ $(\epsilon_-)$ we desorb (adsorb) a brick and 
with probability $1-\epsilon_+$ $ (1 - \epsilon_-)$ we do nothing. In Fig. 2
  we
show for $N = 7, s = 2$ $ (b = 3)$ and $n=2$ some examples of the 
possible configurations of the surface. In this figure we also 
show the corresponding particle configurations in the diffusion 
problem. We can verify  that for arbitrary $s $ (or b), as 
long the growth model is not periodic $(N \ne n b)$ there  exists 
an exact 
one-to-one correspondence between the configurations of particles and 
those of the surface, with the transitions between them described 
by the Hamiltonian \rf{6}. On the other hand, if the growth model 
is periodic $(N = n b)$, there exist $b$  configurations in the  
asymmetric diffusion problem that correspond to a single surface 
configuration (the flat surface). Consequently the Hamiltonian
\rf{6} does not describe exactly the generalized step
model in a finite lattice. 
However as $N$ increases this difference decreases and \rf{6} also 
describes the fluctuations of the growth model.

Finally, in the case where all the molecules has size zero 
a possible growth model is obtained by defining height variable 
$h_i$ $ (i=1,2,...,N)$ at the same position of the molecules in 
the diffusion problem. For a given configuration 
$\lbrace n_1,n_2,...,n_N\rbrace$ with $n_i$ molecules at sites $i$
the height variables in the surface model $(h_{i+1} \geq h_{i})$
satisfy
\beq
\label{f5}
h_i -h_{i-1} = n_i,\;\;\;\; i = 2,...,N,
\eeq
with the boundary  condition
\beq
\label{f6}
h_{N+1} = h_1 + n
\eeq
where $n=\sum n_i$ is the total number of molecules. Bricks of
unity size are desorbed (adsorbed) with transition rates
$\epsilon_+$ $(\epsilon_-)$ if the final configuration satisfies
$h_{i+1}\geq h_i$ $ (i=1,2,\ldots,N-1)$.

\section{ The Bethe ansatz equations}

We present in this section the exact solution of the general quantum chain 
\rf{6}. For simplicity let us consider initially the case where all the 
molecules have the same size  
 $s$ ($0,1 ,\ldots$). In the particular case 
where $s = 1$ we have the standard simple exclusion model whose Bethe 
ansatz solution was obtained by Gwa and Spohn \cite{spohn} and can also be 
obtained after the canonical transformation \rf{12} from the Bethe ansatz 
solution of the XXZ chain with twisted boundary conditions \cite{alcbat1}. 
The exact integrability of the fully asymmetric version of \rf{11} 
($\epsilon_- = 0$), for $s >0$, was verified directly in the master equation 
by Sasamoto and Wadati \cite{sasawada}, and the model with $s =0$ is related to 
the limit $q \rightarrow \infty $ of the $q$-boson hopping model introduced 
by Bogoliubov {\it et al}. \cite{bogo1,bogo2}.

Due to the conservation of particles in the diffusion processes the 
total number of particles are good quantum numbers and we can separate 
the associated Hilbert space into block-disjoint sectors labelled by the 
number $n$ of particles. We therefore consider the eigenvalue equation 
\beq
\label{13}
H |n> = E|n>,
\eeq
where
\beq
\label{14}
|n> = \sum_{\{x\}}f(x_1,\ldots,x_n)|x_1,\ldots,x_n>.
\eeq
Here $x_1,\ldots,x_n$ denotes  the location of particles on the chain and the 
summation extends over all sets $\{x\}$ of the $n$ non-decreasing 
integers satisfying 
\beq
\label{15}
x_i \geq x_{i+1} + s, \quad i=1,\ldots,n-1, \quad s \leq x_n -x_1 \leq N-s.
\eeq
It is important to notice that some of  
 these coordinates may coincide in the case where the particles have 
zero size.

{\it {\bf  n = 1.}}
For one particle on the chain as a consequence of the translational invariance 
of \rf{6} the eigenfunctions are the momentum-$k$ eigenfunctions 
\beq
\label{16}
|1> = \sum_{x=1}^{N} f(x) |x>, \; \; f(x) = e^{ikx},
\eeq
where 
\beq
\label{16-k}
k =\frac{2\pi}{N}l; \quad l=0,1, \ldots,N-1,
\eeq
and energy given by 
\beq
\label{17}
E = e(k) \equiv  -(\epsilon_-e^{ik} + \epsilon_+e^{-ik} -1).
\eeq

{\it {\bf  n =2.}} For two particles on the lattice the eigenvalue equation 
\rf{13} gives  us two distinct relations depending on the relative position of 
the particles. If the two particles are at positions $x_1$ and $x_2$ 
satisfying $x_2 >x_1 +s$ we obtain
\bea
\label{18}
Ef(x_1,x_2) &=& -\epsilon_+f(x_1-1,x_2) - \epsilon_-f(x_1+1,x_2) 
+ 2f(x_1,x_2) \nonumber \\
& & -\epsilon_+f(x_1,x_2-1) - \epsilon_-f(x_1,x_2+1),
\eea
that can be solved by the ansatz
\beq
\label{19}
f(x_1,x_2) = e^{ik_1x_1}e^{ik_2x_2},
\eeq
which gives
\beq
\label{20}
E = e(k_1) + e(k_2).
\eeq
Since the relation \rf{20} is symmetric in $k_1$ and $k_2$ we can write 
a more general solution of \rf{18} as
\beq
\label{21}
f(x_1,x_2) = A_{12}e^{ik_1x_1}e^{ik_2x_2} + A_{21}e^{ik_2x_1}e^{ik_1x_2}
\eeq
with the same energy as in \rf{20}. The second relation happens when 
$x_2 = x_1 + s$. In this case we have 
\beq
\label{22}
Ef(x_1,x_1 +s) = -\epsilon_+f(x_1-1,x_1+s)-\epsilon_-f(x_1,x_1+s+1) 
+ f(x_1,x_1+s) .
\eeq
If we now substitute the ansatz \rf{21} with the energy \rf{20}, the 
constants $A_{12}$ and $A_{21}$, initially arbitrary, should now satisfy 
\beq
\label{23}
\frac{A_{12}}{A_{21}}= - \Bigl(\frac{e^{ik_1}}{e^{ik_2}}\Bigr)
^{s-1}e^{i\Psi_{12}},
\eeq
\beq
\label{24}
e^{i\Psi_{jl}} =\frac{\epsilon_+ + \epsilon_-e^{i(k_j+k_l)} - e^{ik_j}}
{\epsilon_+ + \epsilon_-e^{i(k_j+k_l)} - e^{ik_l}}.
\eeq
The "wave numbers" $k_1$ and $k_2$ are complex in general and 
are fixed due to the cyclic boundary 
condition 
\beq
\label{25}
f(x_2,x_1+N) = f(x_1,x_2),
\eeq
which from \rf{21} give us the equations
\beq
\label{26}
\frac{A_{12}}{A_{21}} = e^{-ik_2N}, \quad \frac{A_{21}}{A_{12}} = e^{-ik_1N}.
\eeq
Equations \rf{23} and \rf{24} give us the Bethe-ansatz equations for 
$n =2$
\beq
\label{27}
e^{ik_jN} = -\prod_{l=1}^2 \Bigl( \frac{e^{ik_j}}{e^{ik_l}}
\Bigr)^{s-1} e^{i\Psi_{j,l}}, 
\; \; j=1,2 \;\;,
\eeq
with energy given by \rf{20}.

{\it {\bf General n.}} The above calculation can be generalized for arbitrary 
values of $n$. The ansatz for the wavefunction becomes 
\beq
\label{28}
f(x_1,\ldots,x_n) = \sum_P A_{P_1,P_2,\ldots,P_n} e^{i(k_{P_1}x_1+\cdots + 
k_{P_n}x_n)},
\eeq
where the sum extends over all permutations $P$ of $1,2,\ldots,n$. If 
$x_{i+1} -x_i >s$ for $i=1,2,\ldots,n$ it is easy to see
that the eigenvalue equation \rf{13} is satisfied by the ansatz \rf{28} 
with energy 
\beq
\label{29}
E = \sum_{j=1}^n e(k_j).
\eeq
If a pair of paticles is at positions $x_i,\; x_{i+1}$, where $x_{i+1} = 
x_i +s$, equation \rf{13} with the ansatz \rf{28} and the relation 
\rf{29} give us 
\beq
\label{30}
\frac{A_{P1,\ldots,P_i,P_{i+1},\ldots,,P_n}}
{A_{P_1,\ldots,P_{i+1},P_i,\ldots,P_n}} = -e^{i(s-1)(k_{P_i} -k_{P_{i+1}})}
e^{i\Psi_{P_i,P_{i+1}}}.
\eeq
Inserting the ansatz  \rf{28} in the boundary condition
\beq
\label{31}
f(x_2,\ldots,x_n,x_1 + N) = f(x_1,\ldots,x_n)
\eeq
we obtain  the additional relation
\beq
\label{32}
A_{P_1,\ldots,P_n} = e^{ik_{P_1}N}A_{P_2,\ldots,P_n,P_1}.
\eeq
If we iterate the relation \rf{30} $n$ times the equation \rf{32} gives us the 
Bethe-ansatz equations
\beq
\label{33}
e^{ik_jN} = (-)^{n-1} \prod_{l=1}^n \Bigl(\frac{e^{ik_j}}{e^{ik_l}}
\Bigr)^{s-1} 
\frac{\epsilon_+ + \epsilon_- e^{i(k_j+k_l)} -e^{ik_j}}
{\epsilon_+ + \epsilon_- e^{i(k_j+k_l)} -e^{ik_l}}
\eeq
for $j=1,2,\ldots,n$. The solutions $\{k_j\}$ of these equations with 
\rf{29} give us the eigenergies of the Hamiltonian \rf{6}. Furthermore, 
it follows from a lattice shifting that the wavefunctions given by the 
ansatz \rf{28} are also eigenfunctions of the momentum operator $\hat{P}$ 
with eigenvalue 
\beq
\label{34}
p = \sum_{j=1}^n k_j \quad \mbox{mod} (2\pi) = \frac{2\pi}{N}l, 
\quad l=0,1,\ldots, 
N-1.
\eeq
In the particular case where $s=1$ equations  \rf{29},\rf{33} and \rf{34} 
recover the results presented in Ref. \cite{spohn}.

Let us now consider the general case where we have $n$ molecules with 
arbitrary given sizes $\{s_1,s_2,\ldots,s_n\}$ ($s_i=0,1,2,\ldots$) and the 
Hamiltonian given by \rf{6}. 
In this case each type of molecule is conserved 
separately. Moreover since in the diffusion processes the particles only 
interchange positions with the vacant sites a given order 
$\{s_1,s_2,\ldots,s_n\}$ of particles remain conserved up to cyclic 
permutations. The wavefunctions can be written as 
\beq
\label{37}
|s_1,s_2,\ldots,s_n> = \sum_{\{c\}} \sum_{\{x\}} 
f^{s_{c_1},\ldots,s_{c_n}}(x_1,\ldots,x_n)|x_1,\ldots,x_n>.
\eeq
Here $f^{s_{c_1},\ldots,s_{c_n}}(x_1,\ldots,x_n)$ is the amplitude of a 
configuration where particles of sizes $s_1,\ldots,s_n$ occupy the 
positions $x_1,\ldots,x_n$, respectively. The summation $\{c\}$ extends over 
all cyclic permutations $\{c_1,\ldots,c_n\}$ of integers $\{1,\ldots,n\}$, 
and the summation $\{x\}$ extends, for a given distribution 
$\{s_{c_1},\ldots,s_{c_n}\}$ of molecules, to increasing integers satisfying 
\beq
\label{38}
x_{i+1} \geq x_i + d_{s_{c_i},s_{c_{i+1}}} \quad  
i =1,\ldots,n-1, \;\; d_{s_{c_n},s_{c_1}} \leq x_n -x_1 \leq 
N- d_{s_{c_n},s_{c_1} } .
\eeq
The ansatz we expect to be valid, that replaces \rf{27} is
\beq
\label{39}
f^{s_1,\ldots,s_n}(x_1,\ldots,x_n) = \sum_P A_{P_1,\ldots,P_n}
^{s_1,\ldots,s_n} e^{i(k_{P_1} + \cdots + k_{P_n})}
\eeq
where $A_{k_{P_1},\ldots,k_{P_n}}^{s_1,\ldots,s_n}$ and $\{k_1,\ldots,k_n\}$ 
are going to be fixed by imposing that \rf{38} with \rf{39} is a solution 
of the eigenvalue equation of the general Hamiltonian \rf{6}.

Let us consider the eigenvectors of \rf{6} with different number of 
particles.

{\it \bf{n = 1.}} For one particle on the chain the ansatz \rf{39} coincides 
with \rf{28} and the wavefunctions and energies are given by \rf{16} 
and \rf{17}, respectively.

{\it \bf {n = 2.}} If both particles are identical $s_1 = s_2=s $,  
we have the same situation considered previously in \rf{18}-\rf{27}.The 
wavefunctions $|s,s>$ are given by \rf{39} with energies given by 
\rf{20} and \rf{27}. However, if the particles are distinct the situation 
is different. If the particles are located at positions $x_1$ and $x_2$, 
with $x_2 -x_1 > d_{s_1,s_2}$ the ansatz \rf{38} is valid with energy 
given by \rf{20} and no restrictions on $\{A_{k_{P_1},k_{P_2}}^{\alpha_i,
\alpha_j}\}$ are necessary. 
If the particles are at the closest distance $x_2 = x_1 + 
d_{s_1,s_2}$ equation \rf{22} is replaced by 
\bea
\label{40}
Ef^{s_1,s_2}(x_1,x_1 + d_{s_1,s_2})& =& -\epsilon_+f^{s_1,s_2}(x_1-1,x_1 +
d_{s_1,s_2}) -\epsilon_-f^{s_1,s_2}(x_1,x_1 +d_{s_1,s_2} +1) \nonumber  \\
& & +f^{s_1,s_2}(x_1,x_1 + d_{s_1,s_2}).
\eea
Inserting in the above equation the ansatz \rf{39} and the energy \rf{20} 
we obtain the relation 
\beq
\label{41}
A_{P_1,P_2}^{s_1,s_2} = -e^{i\Psi_{P_1,P_2}} \
\sum_{s_1',s_2'} S_{s_2',s_1'}^{s_1,s_2}
(k_{P_1} -k_{P_2}) A_{P_2,P_1}^{s_1',s_2'}
\eeq
where $\Psi_{P_1,P_2}$ is given by \rf{24} and the elements of the $S$ 
matrix are given by 
\beq
\label{42}
S_{\gamma,\mu}^{\alpha,\beta}(k) = e^{i(d_{\alpha,\beta}-1)k} \delta_{\alpha,
\mu} \delta_{\beta,\gamma}.
\eeq
 The wave 
numbers $k_1$ and $k_2$ are going to be fixed by the boundary condition 
\beq
\label{43}
f^{s_2,s_1}(x_2,x_1+N) = f^{s_1,s_2}(x_1,x_2),
\eeq
but instead of deriving the Bethe-ansatz equations for $n=2$ let us consider 
the case of general $n$. 

{\it {\bf General n}}. The ansatz \rf{39} applied to the 
case  where two particles are at their closest distance gives us the 
generalization of \rf{41}
\beq
\label{44}
A_{\ldots,P_1,P_2,\ldots}^{\ldots,\alpha,\beta,\ldots} = -e^{i\Psi_{P_1,P_2}} 
\sum_{\alpha',\beta'} S_{\alpha',\beta'}^{\alpha,\beta}
(k_{P_1}-k_{P_2}) A_{\ldots,P_2,P_1,\ldots}^{\ldots,\beta',\alpha',\ldots},
\eeq
with $S$ given by \rf{42}. Successive applications of this equation give us in 
general different relations between the amplitudes. For example $A_{\ldots,
k_1,k_2,k_3,\ldots}^{\ldots,\alpha,\beta,\gamma,\ldots}$ relate to 
$A_{\ldots,k_3,k_2,k_1,\ldots}^{\ldots,\gamma,\beta,\alpha,\ldots}$ by 
performing the permutations $\alpha\beta\gamma \rightarrow \beta\alpha\gamma 
\rightarrow \beta\gamma\alpha \rightarrow \gamma\beta\alpha$ or 
 $\alpha\beta\gamma \rightarrow \alpha\gamma\beta  
\rightarrow \gamma\alpha\beta \rightarrow \gamma\beta\alpha$, and consequently 
the $S$-matrix should satisfy the Yang-Baxter \cite{yang,baxter} equation
\bea
\label{45}
\sum_{\gamma,\gamma',\gamma''} S_{\gamma,\gamma'}^{\alpha,\alpha'}(k_1-k_2)
S_{\beta,\gamma''}^{\gamma,\alpha''}(k_1-k_3)& &S_{\beta',\beta''}^
{\gamma',\gamma''}(k_2-k_3) = \nonumber \\
& & \sum_{\gamma,\gamma',\gamma''}
 S_{\gamma',\gamma''}^{\alpha',\alpha''}(k_2-k_3)
S_{\gamma,\beta''}^{\alpha,\gamma''}(k_1-k_3)S_{\beta,\beta'}^
{\gamma,\gamma'}(k_1-k_2). 
\eea
 Actually the relation \rf{45} 
is a necessary and sufficient condition \cite{yang,baxter} 
to obtain non-trivial 
solution for the amplitudes in \rf{44}. The validity of \rf{45} can easily 
be verified for the diagonal $S$-matrix \rf{42} in the problem we are 
considering.

The boundary condition
\beq
\label{46}
f^{s_1,\ldots,s_n}(x_1,\ldots,x_n) = f^{s_2,\ldots,s_n,s_1}(x_2,\ldots,x_n,x_1)
\eeq
implies the relation between the amplitudes 
\beq
\label{47}
A_{P_1,\ldots,P_n}^{s_1,\ldots,s_n} = e^{ik_{P_1}N} 
A_{P_2,\ldots,P_n,P_1}^{s_2,\ldots,s_n,s_1}.
\eeq
If we now apply relation \rf{44} $n$ times we can obtain a relation between 
the amplitudes with same momenta, i. e., 
\bea
\label{48}
& & A_{P_1,\ldots,P_n}^{s_1,\ldots,s_n} = (-)^{n-1} e^{i\sum_{l=2}^n
\Psi_{P_l,P_1}} e^{ik_{P_1}N} \nonumber \\
& & \sum_{\{s_1',\ldots,s_n'\}} \sum_{\{s_1'',\ldots,s_n''\}} 
S_{s_1',s_1''}^{s_1,s_2''}(k_{P_1}-k_{P_1}) 
S_{s_2',s_2''}^{s_2,s_3''}(k_{P_2}-k_{P_1}) \cdots 
S_{s_n',s_n''}^{s_n,s_1''}(k_{P_n}-k_{P_1}) A_{P_1,\ldots,P_n}^
{s_1',\ldots,s_n'},
\eea
where we introduced the extra sum 
\beq
\label{49}
1 = \sum_{s_2'',s_1''} \delta_{s_2'',s_1'} \delta_{s_1,s_1''} =
 \sum_{s_2'',s_1''} S_{s_1',s_1''}^{s_1,s_2''}(k_{P_1}-k_{P_1}).
\eeq
In order to fix the values of $\{k_j\}$ we should then find the eigenvalues 
$\Lambda (k)$ of the matrix
\beq
\label{50}
<\{s\}|T(k)|\{s'\}> = \sum_{\{s_1'',\ldots,s_n''\}} \prod_{l=1}^n 
S_{s_l',s_l''}^{s_l,s_{l+1}''}(k_{P_l} -k),
\eeq
with $s_{n+1}'' = s_1''$. 
 We identify $T(k)$ as the transfer matrix of a inhomogeneous vertex model, 
with inhomogeneities $(k_{P_l} - k)$, in a periodic lattice. 
The Boltzmann weights of the vertex models are given by $S_{\gamma \delta}
^{\alpha \beta}$ and the number of distinct vertices depends on the particular 
configuration (type of order) of molecules in our diffusive system. 
Using \rf{42} we can see that there exists only one non-zero element 
for each line, i. e., $<s_1,\ldots,s_n|T|s_2,\ldots,s_n,s_1>$.

In order to calculate the eigenvalues $\Lambda(k)$ we apply the transfer 
matrix $r$ times in the state $A^{s_1,\ldots,s_n}$, where $r$ is the 
minimum number of cyclic rotations of $\{s_1,\ldots,s_n\}$ where the 
configuration repeats the initial one. We may show  that 
\beq
\label{51}
T^rA_{P_1,\ldots,P_n}^{\alpha_1,\ldots,\alpha_n} = \Lambda^r(k) A_{P_1,\ldots,P_n}^{\alpha_1,\ldots,\alpha_n}.
\eeq
Also  it is easy to compute
\beq
\label{52}
\Lambda^r(k) =\exp \Bigl[ i\frac{r}{n} (\sum_{l=1}^n d_{s_l,s_{l+1}}-1)
\sum_{l=1}^n (k_l-k) \Bigr].
\eeq
Finally, substituting $\Lambda(k_{P_1})$ in \rf{48} we obtain the 
Bethe-ansatz equations
\beq
\label{53}
e^{ik_jN} = e^{i\frac{2\pi}{r}m} (-1)^{n-1} \prod_{l=1}^n
e^{i(k_j-k_l)(\tilde s-1)} \frac{\epsilon_+ + \epsilon_- e^{i(k_j+k_l)} - 
e^{ik_j}}{\epsilon_+ + \epsilon_-e^{i(k_j+k_l)} - e^{ik_l}},
\eeq
where $j=1,\ldots,n$;  $m=0,1,\ldots,r-1$ and 
\beq
\label{54}
\tilde s = \frac{1}{n} \sum_{l=1}^n d_{s_l,s_{l+1}}
\eeq
plays the role of an average molecule size  of the particular
 configuration $\{s\}$ of molecules. As we can see, by comparing \rf{53} with 
\rf{33} the extra phase $\exp(i2\pi m/r)$ ($m =0,1,\ldots,r-1$) gives $r$ times 
more solutions of \rf{53} than in \rf{33}. This indeed should be the case 
since the Hilbert space associated to the Hamiltonian \rf{6} of a given 
distribution of particles of sizes $\{s_1,\ldots,s_n\}$, due to the 
distinguibility of the particles, is $r$ times bigger than that of the 
Hamiltonian \rf{6} when $s_1=s_2=\cdots=s_n$. It is interesting to observe 
that $\tilde s$ can take any non-negative 
rational number by choosing appropriately 
$\{s_1,\ldots,s_n\}$. Also many distinct distributions of molecule's sizes 
with the same effective $\tilde s$ will have the same eigenenergies.

In an appendix we explore our Bethe-ansatz solution to obtain the relationship 
between the eigenvalues and eigenvectors of the Hamiltonian \rf{6} with 
different  distributions of molecule's sizes. 

\section{The critical exponent z}

In this section we calculate the dynamical critical exponent $z$ for the 
stochastic models presented in Sec. 2. This calculation is done by exploiting 
its connection with the mass gap $G_N$ of the corresponding Hamiltonian,
\beq
\label{55}
\mbox{Re}(G_N) \sim N^{-z}.
\eeq

A calculation for arbitrary values of $\epsilon_+,\epsilon_-$ and density 
$n/N$ can be done systematically by using the method presented in \cite{kim}. 
However, since universality arguments indicate that $z$ should be independent  
on the particular values of $\epsilon_+, \epsilon_-$ and $n$,  as long as 
$\epsilon_+\neq \epsilon_-$, we are going to restrict ourselves, like in
\cite{spohn}, to the simplest case where $\epsilon_- =0, \epsilon_+ =1$. A 
general discussion for the other cases, which does not change our results 
is given  at the end of the appendix. Defining the variables
\beq
\label{56}
z_j = 2e^{-ik_j} -1
\eeq
the energies \rf{29} and momenta \rf{34} are given by 
\beq
\label{57}
E =  \sum_{j=1}^n (1 - z_j)/2,
\eeq
\beq
\label{58}
e^{-iP} = \prod_{j=1}^n (1 + z_j)/2,
\eeq
respectively. The $\{z_j\}$ variables should satisfy the Bethe-ansatz 
equations \rf{53}
\beq
\label{59}
(1+z_j)^{N-n\tilde s} (1 - z_j)^n = -2^N e^{-i\frac{2\pi}{r} m} 
\prod_{l=1}^n \frac{z_l -1}{(z_l+1)^{\tilde s}} 
\eeq
where $j=1,\ldots,n$, $m=0,1,\ldots,r-1$. It is interesting to note that 
these equations are simpler than the usual Bethe-ansatz equations appearing 
in other exact integrable systems since the right-hand side of \rf{59}  is 
 independent of  the particular value of $j$. These equations are even 
simpler  for the special "half-filled" density 
$\rho = n/N = 1/(1+\tilde s)$, i. e., 
\bea
\label{60}
(1 - z^2)^n = Y \nonumber \\
Y = -2^{n(1 + \tilde s)} \prod_{l=1}^n \frac{z_l -1}{(z_l+1)^{\tilde s}}.
\eea
If we parametrize $Y = -a^n e^{in \theta }$, with $a \geq 0$ and 
$\theta \in  (-\frac{\pi}{n},\frac{\pi}{n})$, the $2n$ roots $z_j$ are 
given by
\bea
\label{61}
z_j &=& (1 - y_j)^{\frac{1}{2}}, 
\quad z_{j+n} = - z_j;  \nonumber \\
y_j &=& a e^{i\theta} e^{i2\pi(j-\frac{1}{2})/n};  \quad j =1,\ldots,n.
\eea
For a given choice $\{z_j\}$ of the above set and a given value of $m$ 
($0,1,\ldots,r-1$) we have only two unknowns, $a$ and $\theta$, which are 
obtained from the  equation
\beq
\label{62}
(ae^{i\theta})^n = e^{i\frac{2\pi}{r}m} 2^{(\tilde s +1)n} 
\prod_{l=1}^n \frac{z_{l(j)} -1}{[z_{l(j)} +1]^{\tilde s}}.
\eeq
We have solved numerically the above equations for several values of 
$\tilde s, m, r$ and $n$. The ground-state energy $E=0$ is 
obtained by choosing  $m=0$ in \rf{62}, and is given by the 
configuration 
\beq
\label{63}
C_0 = \{z_1,z_2,\ldots,z_n\},
\eeq
with  $a = \theta = 0$. In order to find the first excited state we should 
consider different choices of $\{z_j\}$ and different values of $m$. 
Since $z_j + z_{n+j} = 0$ the energy increases as we take, in the 
configurations $\{z_j\}$, values of $z_j$ where $n<j\leq 2n$. 
Therefore configurations $\{z\}$ associated with  
low energies are
\beq
\label{64}
C_1 = \{z_1,z_2,\ldots,z_{n-1},z_{n+1}\}
\eeq
and
\beq
\label{65}
C_{-1} = \{z_{2},\ldots,z_{n-1},z_n,z_{2n}\}.
\eeq
These configurations, from \rf{58}, correspond to states with momentum 
$-\frac{2\pi}{N}$ and $\frac{2\pi}{N}$, respectively. Our numerical 
results show that the energy corresponding to the configuration $C_0$ with 
$m\neq 0$ behaves for large $N$ as 
\beq
\label{66}
E_{C_0,m} \sim 
\frac{a}{n^{\frac{1}{2}}} - i\frac{\pi}{r(\tilde s +1)}m  \quad 
m = 1, 2, \ldots \;\;.
\eeq
On the other hand, the configuration $C_1$ or $C_{-1}$, for sufficiently 
large values of $N$ produces the lowest energy when $m=0$, independently of 
$\tilde s, r$ and behaves as 
\beq
\label{67}
E_{C_{\pm 1},0} \sim \frac{a_0}{n^z} + i\frac{b_0}{n^\gamma},  \; \; 
z = \frac{3}{2}, \;\; \gamma =1,
\eeq
where $a_0$ and $b_0$ are constants. The energies for different values of $m$ 
but with configurations $C_{\pm 1}$ also behave similarly as  \rf{67}. 
Comparing \rf{67} with \rf{66} we see that the gap is given by 
$E_{C_{\pm 1},0}$ and is real only for the special case $\tilde s =1$, treated in \cite{spohn}. 
The values of $a$ and $\theta$ that correspond to the first excited state 
behaves asymptotically as 
\beq
\label{68}
a = 1 + \frac{\beta}{n} + o(n^{-1}), \;\; \theta = \pm \alpha\frac{ 
(\tilde s -1)} {n^{\frac{3}{2}}} + o(n^{-\frac{3}{2}}),
\eeq
where $\beta$ and $\alpha$ are constants. 
In order to illustrate our numerical results we show in Table 1 the 
finite-size estimates for the amplitudes $a_0, b_0$ and the exponents $z$ and 
$\gamma$ defined in equation \rf{67}.

Accepting the behavior \rf{68} for the values of $a$ and $\theta$ for the 
first excited state we also used the same procedure as in Gwa and Spohn 
\cite{spohn} in order to show analitically that $z = \frac{3}{2}$,  
$a_0= 2.301\;345\;96 \ldots$, independently of the  value of $\tilde s$ and 
$b_0 = \pi(\tilde s -1)/(2(\tilde s + 1))$. In the last line of Table 1 we 
show the exact results obtained analitically. 

These results indicate that all these models with arbitrary mixture if 
molecules of different sizes, as well as the corresponding generalized 
growth models exhibit a universal behavior with a KPZ-type of dynamical 
behavior. In the appendix we show that for general values of $\epsilon_+,
\epsilon_-$ and $n$ the wave functions of \rf{6} for arbitrary distributions 
of molecule's sizes, are exactly related. This implies that conditional 
probabilities and correlation functions for arbitrary distribution $\{s\}$ 
are exactly related to the simple exclusion problem $\{s_1 = s_2 = \cdots 
s_n =1\}$. The eigenvalues of these models are exactly related in the case 
of free boundaries. In the case we have a periodic lattice the eigenvalues 
of $H^{\{s\}}$ are exactly related to the asymmetric XXZ chain with 
twisted boundary condition $\phi$ proportional to the momentum of the 
first excited state. Since the momentum of this state is $P =2\pi/N$ the 
effect of the twisted angle should not affect the leading order in the mass 
gap calculations. This implies that for arbitrary values of $\epsilon_+, 
\epsilon_-$ and density $n$ the leading order results of the real part 
of the gap are the same as those calculated systematically in \cite{kim}.


\section{Concluding remarks}

We have solved exactly a general asymmetric diffusion problem where the 
particles may have distinct and arbitrary integer sizes. 
We also show in Sec. 2 that these diffusion 
models are related to generalized growth models.  Since through diffusion 
the particles do not interchange positions, a 
given order $\{s_1,\ldots,s_n\}$ of the distribution of molecule's sizes on 
the lattice is fixed, up to cyclic permutations. A parameter which is 
proportional to the excluded volume for the particles is 
the average size of the molecules $\tilde s$ given 
by \rf{54}. In the case of the simple exclusion problem all the molecules 
have the same unity size $s_1 = s_2 =\cdots =s_n = 1$, which gives 
$\tilde s = 1$. On the other hand, if all the particles has no size  we have 
$\tilde s =0$ and there is no excluded volume. By choosing suitable 
distribution $\{s\}$ of molecule's sizes we may change $\tilde s$ almost 
continuously in the bulk limit $N \rightarrow \infty$.  Exploiting the 
connection between the dyanmical critical exponent $z$ and the mass gap 
of the related quantum chain we obtained $z = \frac{3}{2}$ for all the 
models, independently of the parameter $\tilde s$ measuring the excluded 
volume. This implies that, at least in one dimension, the excluded 
volume effect is irrelevant for dynamical systems in the KPZ universality class.

We also show (see the appendix) that the wave functions of the models 
with arbitrary distribution of molecule's sizes can be related to those of a 
simple asymmetric exclusion problem, in a distinct lattice size. This 
implies that conditional probabilities and correlation functions of these 
models are exactly related.
\acknowledgements {
This work was supported in part by Conselho Nacional de Desenvolvimento 
Cient\'{\i}fico e Tecnol\'ogico - CNPq - Brasil, by  FINEP - Brasil and 
by the Russian Foundation of Fundamental Investigation ( Grant 99-02-17646). }
\appendix
\section{ Exact spectral relations between the models  with distinct  
molecule's distribution}
In this appendix we are going to show how the eigenvalues and eigenvectors of 
the general Hamiltonian \rf{6} with different distribution of molecules are 
related to those of the simple asymmetric exclusion Hamiltonian \rf{1}. Let 
us consider initially the case of free boundary conditions. In this case 
we have to specify, for a given lattice size $N$, the occupation at the 
border of the lattice, i. e., the minimal $x_I$ and maximum $x_F$ 
coordinates  we may put a molecule of size $s$. As an extension of \rf{1-1} 
we define these coordinates, for the model with a distribution 
$\{s_1,s_2,\ldots,s_n\}$ of molecules as
\beq
\label{a1}
x_I = d_{s_1,0} + \delta_{s_1,0} = \mbox{Int}[(s_1+1)/2] + \delta_{s_1,0} 
, \;\; x_F = N - (d_{s_n,0} -1) - \delta_{s_n,0}.
\eeq
If $|x_1,\ldots,x_n>$ are the vectors corresponding to the coordinates of 
the particles of sizes $\{s_1,\ldots,s_n\}$ the application of the 
Hamiltonian \rf{6} with free 
ends in a given vector can be written as
\bea
\label{a2}
H_{\{s\}}^F& & |x_1,\ldots,x_n> = \nonumber \\ 
 & & -\sum_{i=1}^{n-1} \epsilon_+ \theta(x_{i+1} -x_i -d_{s_i,s_{i+1}}) 
[|x_1,\ldots,x_{i-1},x_i+1,x_{i+1},\ldots,x_n> - |x_1,\ldots,x_n>] 
\nonumber \\
& & -\sum_{i=2}^{n} \epsilon_- \theta(x_{i}-x_{i-1}-d_{s_{i-1},s_i}) 
[x_1,\ldots,x_{i-1},x_i-1,x_{i+1},\ldots,x_n> - |x_1,\ldots,x_n>] \nonumber \\ 
& & -\epsilon_+\theta(x_F -x_n)[|x_1,\ldots,x_{n-1},x_n+1> - 
|x_1,\ldots,x_n>] \nonumber \\
& & -\epsilon_-\theta(x_1-x_I)[|x_1-1,x_2,\ldots,x_n> -
|x_1,\ldots,x_n>],
\eea
where $\theta(x) = 0$ for $x \leq 0$ and $\theta(x) = 1$ for $x >0$, is the 
standard step function.

If we now make the following change of coordinates 
\beq
\label{a3}
x_i' = x_i - x_I -\sum_{j=1}^{i-1} d_{s_j,s_{j+1}} + i
\eeq
we can rewrite \rf{a1} as
\bea
\label{a4}
H_{\{s\}}^F & & |x_1',\ldots,x_n'>  = \nonumber \\
& & -\sum_{i=1}^{n-1} \epsilon_+\theta(x_{i+1}' -x_i' -1) 
[|x_1',\ldots,x_{i-1}',x_i'+1,\ldots,x_n'> - |x_1',\ldots,x_n'>] \nonumber \\
& &  -\sum_{i=2}^{n} \epsilon_-\theta(x_{i}' -x_{i-1}' -1) 
[|x_1',\ldots,x_{i-1}',x_i'-1,x_{i+1}',\ldots,x_n'> - |x_1',\ldots,x_n'>] 
\nonumber \\
& & -\epsilon_+\theta(x_F'- x_n')[|x_1',\ldots,x_{n-1}',x_n' +1> - 
|x_1',\ldots,x_n'>] \nonumber \\ 
& & -\epsilon_-\theta(x_1'- x_I')[|x_1'-1,x_2',\ldots,x_n'> - 
|x_1',\ldots,x_n'>],
\eea
where now $x_I' =1$ and $x_F' = x_F - x_I - \sum_{j=1}^{n-1} d_{s_j,s_{j+1}} + 
n \equiv  N'$. But this is exactly the application of the 
simple asymmetric exclusion Hamiltonian $H_{\{s_1=\cdots=s_n=1\}}^F$   in a 
lattice size $N'$. Consequently for free ends there exists a one-to-one 
correspondence between the eigenvalues and eigenvectors of our general 
Hamiltonian \rf{6} with arbitrary distribution of particle's sizes with that 
of the standard simple diffusion problem, in a lattice size which depends 
on the volume excluded due to the molecule's sizes, i. e.,
\bea
\label{a5}
H_{\{s\}}^F (n,N) = H_{\{s_1=s_2=\cdots=s_n=1\}}^F(n,N')  \\
N' = N - \sum_{j=1}^{n-1} d_{s_j,s_{j+1}} -d_{s_1,0} - 
d_{s_n,0} + n +1 -\delta_{s_1,0} -\delta_{s_n,0}.
\eea
The Bethe ansatz solution of the XXZ chain with surface fields given in 
\cite{alcbat2}, after the canonical transformation \rf{10} can be easily 
exported for our general model \rf{6} with free ends. As observed in 
\cite{alcrit2} the simple asymmetric diffusion Hamiltonian \rf{1}, with 
free boundaries, has a quantum $U_qSU(2)$ symmetry with $q=\sqrt{\frac{
\epsilon_+}{\epsilon_-}}$. This symmetry implies an exact form for the 
ground-state wavefunction. Using this wavefunction in the relation 
\rf{a3} we obtain the ground state wavefunction for the general 
Hamiltonian \rf{6} with free ends
\beq
\label{a6}
\Psi_0^{\{s\}} = \sum_{\{x\}} \prod_{i=1}^n 
\Bigl(\frac{\epsilon_+}{\epsilon_-}\Bigr)^{x_i +i - \sum_{j=1}^{i-1}d_{s_j,s_{j+1}}}) 
|x_1,x_2,\ldots,x_n>.
\eeq

Let us now consider the case of the Hamiltonian \rf{6} with twisted boundary 
conditions specifyed by the angle $\phi$, which in general is a complex 
number,
\beq
\label{a7}
E_{N+1}^{\beta,0} = e^{i\phi} E_1^{\beta,0}, \;\; E_{N+1}^{0,\beta} = 
e^{-i\phi} E_1^{0,\beta}, \;\; \beta \neq 0 \\
E_{n+1}^{0,0} = E_1^{0,0}, \;\; E_{N+1}^{\beta,\beta} = E_1^{\beta,\beta}.
\eeq
The periodic case treated in section (3) corresponds to 
the case where $\phi =0$. The 
Bethe-ansatz equations for these boundary conditions can be obtained 
by changing \rf{46} and \rf{47}, and are  given by
\beq
\label{a8}
e^{ik_jN'}e^{i\tilde \phi_m} =  (-1)^{n-1}\prod_{l=1}^n
\frac{\epsilon_+ + \epsilon_- e^{i(k_j+k_l)} - 
e^{ik_j}}{\epsilon_+ + \epsilon_- e^{i(k_j+k_l)} - e^{ik_l}},
\eeq
where 
\bea
\label{a9}
N' = N -n(\tilde s -1), \;\; \tilde \phi_m = \phi + P(\tilde s -1) + 
\frac{2\pi}{r}m, \;\; m=0,1,\ldots,r-1, \\
P = \sum_j k_j \;\;\ \mbox{mod}(2\pi) = \frac{2\pi}{N}l, \;\; l =0,1,\ldots,N-1,
\eea
is the momentum and $r,\tilde s$ are defined in \rf{51}-\rf{54}. These 
equations give us the following equivalence between the eigenspectra of the 
general Hamiltonian \rf{6} with boundary condition $\phi$, in the 
eigensector with $n$ particles and a given momenta $P$
\beq
\label{a10}
H_{\{s_1,\ldots,s_n\}}^{\phi}(N,P,n) 
 = \sum_{m=0}^{r-1} H_{\{s_1 =s_2=\cdots=s_n=1\}} 
^{\tilde \phi_m} (N',P,n),
\eeq
where in the right-hand side we have the eigenspectra of the asymmetric 
simple exclusion Hamiltonian \rf{1} with twisted boundary condition 
$\tilde \phi_m$. In the right-hand  side of the above 
equation we have also to add several 
eigenspectra, depending on the value of $r$ (see Sec. 3), and this is due 
to the distinguibility of the particles in the Hamiltonian of the left-hand 
 side 
of the equation. 

Our Bethe solutions presented in Sec. 3 also give us a connection between the 
wavefunctions of the models. The eigenstates related by \rf{a10}, apart 
from an overall normalization  satisfy
\beq
\label{a11}
\Psi_{\{s_1,\ldots,s_n\}}^{\{k_1,\ldots,k_n\}}(x_1,x_2,\ldots,x_n)  =  
\Psi_{\{s_1+s_2=\cdots=s_n=1\}}^{\{k_1,\ldots,k_n\}}(x_1',x_2',\ldots,x_n')
\eeq
where
\beq
\label{a12}
x_1' = x_1, \;\; x_i' = x_i + i -1 - \sum_{j=1}^{i-1}d_{s_j,s_{j+1}}, \;\;\; 
i =2,\ldots,n.
\eeq
The results \rf{a5} and \rf{a11}-\rf{a12} imply that any calculation 
involving only wavefunctions can be straigtforwardly translated for arbitrary 
distribution $\{s\}$ of molecule's sizes. An example of this is the 
$n$-particle Green's function $P^{\{s\}}(x_1,\ldots,x_n;t|y_1,\ldots,y_n;0)$ 
which gives the probability of finding particles of size $s_i$, initially 
($t=0$) at  $y_i$ and at time $t$ at $x_i$ ($i=1,\ldots,n$). These 
Green's functions for the different models satisfy
\beq
\label{a13}
P^{\{s_1,\ldots,s_n\}}(x_1,\ldots,x_n;t,y_1,\ldots,y_n;0) = 
P^{\{s_1,=\cdots = s_n=1\}}(x_1',\ldots,x_n';t,y_1',\ldots,y_n';0),  
\eeq
where $x_i'$ and $y_i'$ are related to $x_i$ and $y_i$ as in \rf{a3}. The 
above result generalizes that obtained by Sasamoto and Wadati \cite{sasawada}
 for the case we have a fully asymmetric model 
($\epsilon_- = 0$) and molecules of identical 
sizes $s_1 = s_2 =\cdots=s_n=s$.

Calculation involving eigenvalues, like the calculation of the exponent $z$ 
we did in Sec. 4 for the fully asymmetric model $(\epsilon_- = 0)$ should 
be translated 
with care among the different models.  The Bethe-ansatz equations  
\rf{a8} tell us that the eigenvalues of our general model \rf{6} in a periodic 
lattice of size $N$ ($\phi=0$) are the same as those of the simple exclusion 
Hamiltonian \rf{1} in a lattice of size $N' = N - n(\tilde s-1)$ and 
twisted boundary condition $\tilde \phi_m = P(\tilde s-1) + \frac{2\pi}{r}m$, 
 $m = 0,1,\ldots, r-1$. However the eigenvalues of the simple asymmetric 
exclusion Hamiltonian \rf{1} depend on the boundary condition. Our results of 
Sec. 3 although valid only for $\epsilon_- =0$ indicate that the effect of the 
boundary angle in the finite-size corrections is   of higger order 
than the leading corrections for the 
first excited state, since in this case $\tilde \phi_m = \tilde \phi_0 = 
2\pi(\tilde s-1)/N$. This imply that for arbitrary distribution of molecules 
and arbitrary values of $\epsilon_+,\epsilon_-$ and densities we can use the 
results obtained in \cite{kim}, for the leading order of the real part of the 
mass gap, which gives a universal dynamical critical exponent 
$z = \frac{3}{2}$ of KPZ type.

\newpage 
%

%
%


\newpage
\Large
\begin{center}
Figure Captions
\end{center}
\normalsize
\vspace{1cm}

\noindent Figure 1 - Example of configurations of molecules with distinct 
sizes in the asymmetric diffusion problem. These examples correspond to $n=5$ 
particles in a lattice with size $N=6$. The sizes $\{s_1,s_2,\ldots,s_5\}$ 
are shown in figures (a)-(c).

\noindent Figure 2 - Examples of configurations in the growth model with 
$N=7$, $b =3$ and $n =2$. The corresponding configurations of particles 
in the asymmetric diffusion problem (Eq. \rf{6}) are also shown. 
The configurations (b) and (d) are obtained by adsorbing a brick (size $b=3$) 
or by moving a particles to the left in configuration (a), and 
configuration (c) and (e) are obtained by desorbing a brick, or by moving a 
particle to the right in configuration (a).

\vspace{2cm}
\begin{center}
\Large
Table Caption
\end{center}
\normalsize
\vspace{1cm}
\noindent Table 1 -  Examples of finite-size estimates for the 
amplitudes $a_0$, $b_0$ and the exponents $z$ and $\gamma$ in \rf{67}. These 
estimates correspond to the cases $\tilde s = 0.25$ and $\tilde s= 2.5$. The 
calculated  analitical results in the $n \rightarrow \infty$ are also shown 
in the last line of this table.

\newpage

\vspace{2cm}
\Large
\begin{center}
Table 3
\normalsize
\vspace{1cm}
 
\begin{tabular}{|c|c|c|c|c|c|c|c|c|}  \hline
 $n$ & \multicolumn{4} {c|} {$\tilde s =0.25$} &
  \multicolumn{4} {c|} {$\tilde s =2.5$} \\ \hline 
 & $a_0$ & $b_0$ & $z$ & $\gamma$ &
  $a_0$ & $b_0$ & $z$ & $\gamma$  \\ \hline
10 & 2.522601 &  -0.916488 & 1.586388 &  0.934332 &
 2.495387  & 0.654527 & 1.576255 & 0.934156 \\ \hline
50 &  2.351517 &  -0.941304 &  1.521414  &  0.997428  & 
2.345566 & 0.672351 & 1.518526 & 0.997403 \\ \hline
100 &   2.326504 &  -0.942187 & 1.510838 & 0.999374 & 
2.323540 & 0.672989 & 1.509564 & 0.999367 \\ \hline
150 &   2.318133 &  -0.942349 &  1.507253 & 0.999724 &
2.316160 & 0.673106  & 1.506402 & 0.999721 \\ \hline 
200  &  2.313942 &   -0.942406 &  1.505450 & 0.999846 &
 2.312463 & 0.673146 & 1.504811 & 0.999844 \\ \hline 
250  &  2.311425 &   -0.942432  & 1.504365  & 0.999902 & 
2.310243 & 0.673165 & 1.503853 & 0.999900 \\ \hline 
300  &  2.309747 &   -0.942446 & 1.503640  & 0.999932 & 
2.308737 & 0.673175 & 1.503189 & 0.999957 \\ \hline 
 400 &   2.307648 &  -0.942460 & 1.502733 &  0.999962 &
       &   &   & \\ \hline 
500 &   2.306388  &   -0.942466 &  1.502186 &  0.999976 & 
       &   &   & \\ \hline 
800 &   2.304498 &   -0.942473  & 1.501364 &   0.999988 & 
       &   &   & \\ \hline \hline 
EXACT & 2.301346 & -0.942478 & 1.5 & 1 & 2.301346 & 0.673198 & 1.5 & 1 \\ 
\hline 
\end{tabular}
\end{center}

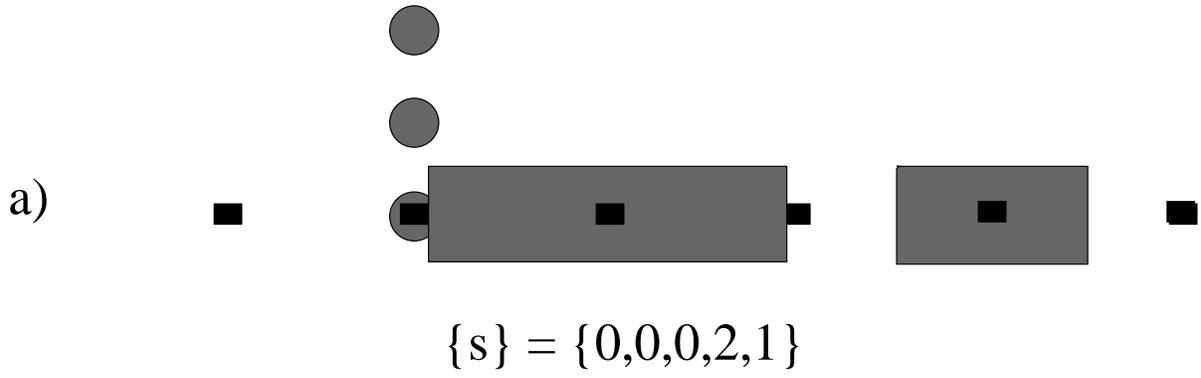
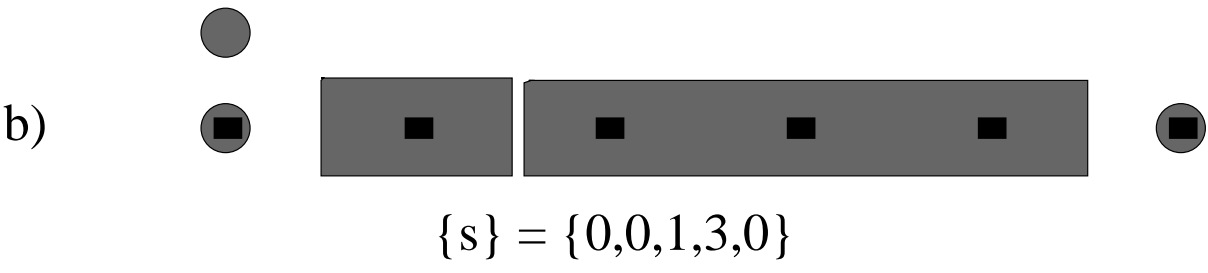
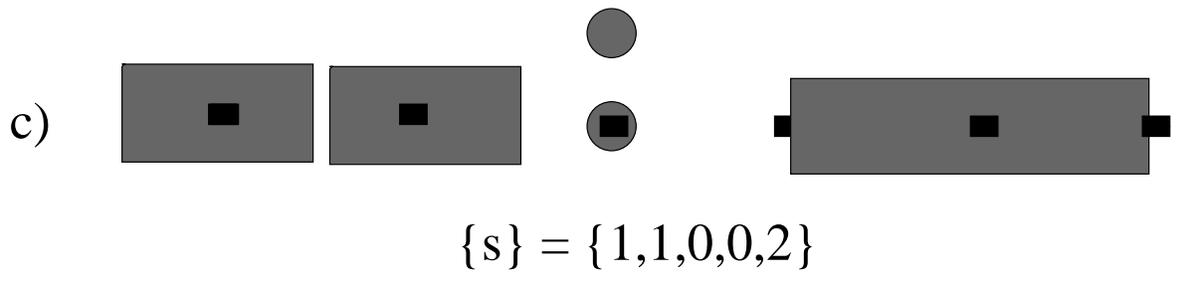
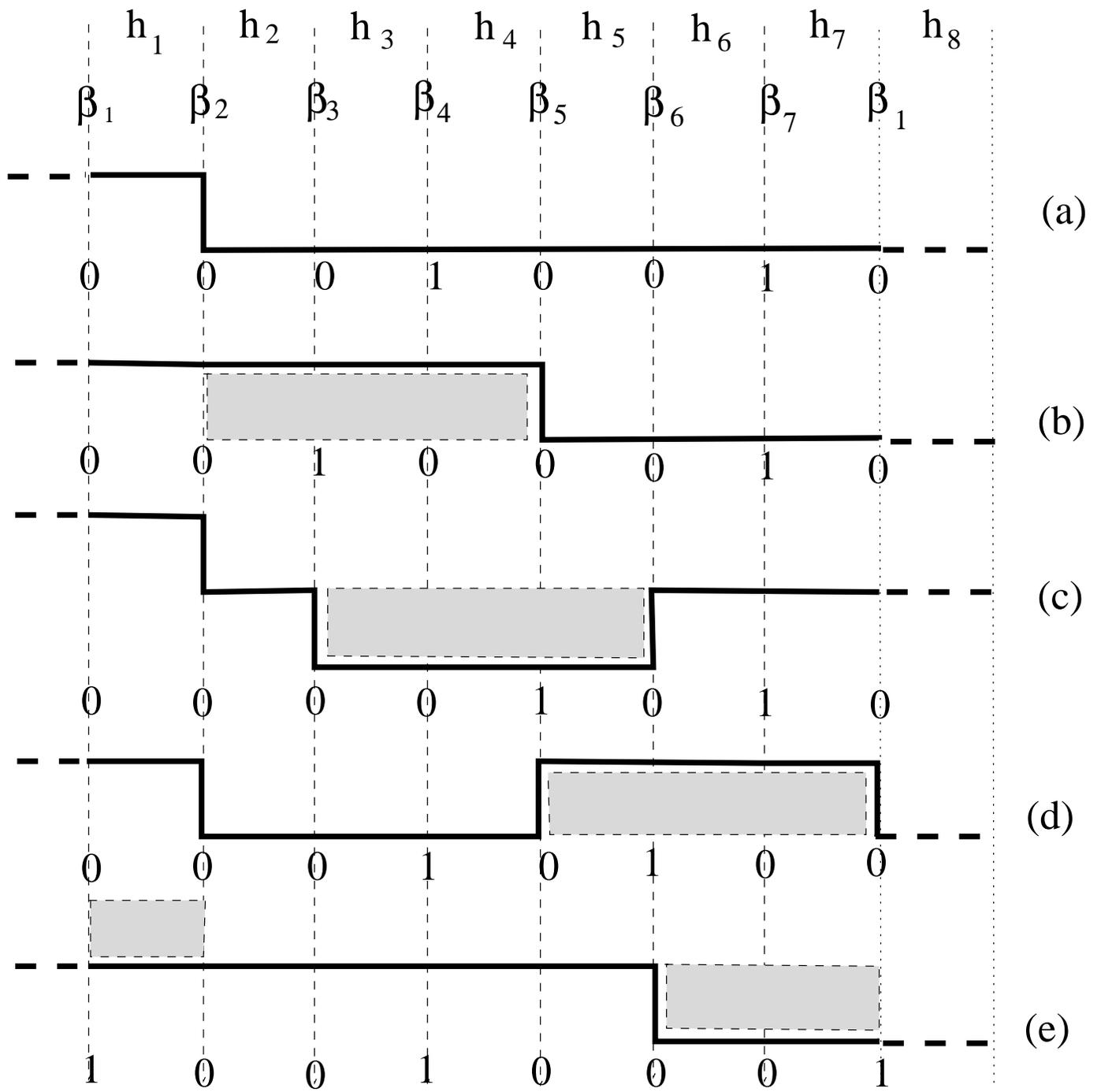
\end{document}